\documentstyle[twoside]{article}

\catcode`\@=11
\long\def\@makefntext#1{
\protect\noindent \hbox to 3.2pt {\hskip-.9pt
$^{{\eightrm\@thefnmark}}$\hfil}#1\hfill}               

\def\thefootnote{\fnsymbol{footnote}}
\def\@makefnmark{\hbox to 0pt{$^{\@thefnmark}$\hss}}    

\def\ps@myheadings{\let\@mkboth\@gobbletwo
\def\@oddhead{\hbox{}
\rightmark\hfil\eightrm\thepage}
\def\@oddfoot{}\def\@evenhead{\eightrm\thepage\hfil
\leftmark\hbox{}}\def\@evenfoot{}
\def\sectionmark##1{}\def\subsectionmark##1{}}

\oddsidemargin=\evensidemargin
\renewcommand{\thefootnote}{\fnsymbol{footnote}}

\newcounter{sectionc}\newcounter{subsectionc}\newcounter{subsubsectionc}
\renewcommand{\section}[1] {\vspace{12pt}\addtocounter{sectionc}{1}
\setcounter{subsectionc}{0}\setcounter{subsubsectionc}{0}\noindent
        {\tenbf\thesectionc. #1}\par\vspace{5pt}}
\renewcommand{\subsection}[1] {\vspace{12pt}\addtocounter{subsectionc}{1}
        \setcounter{subsubsectionc}{0}\noindent
        {\bf\thesectionc.\thesubsectionc. {\kern1pt \bfit #1}}\par\vspace{5pt}}
\renewcommand{\subsubsection}[1] {\vspace{12pt}\addtocounter{subsubsectionc}{1}
        \noindent{\tenrm\thesectionc.\thesubsectionc.\thesubsubsectionc.
        {\kern1pt \tenit #1}}\par\vspace{5pt}}
\newcommand{\nonumsection}[1] {\vspace{12pt}\noindent{\tenbf #1}
        \par\vspace{5pt}}

\newcounter{appendixc}
\newcounter{subappendixc}[appendixc]
\newcounter{subsubappendixc}[subappendixc]
\renewcommand{\thesubappendixc}{\Alph{appendixc}.\arabic{subappendixc}}
\renewcommand{\thesubsubappendixc}
        {\Alph{appendixc}.\arabic{subappendixc}.\arabic{subsubappendixc}}

\renewcommand{\appendix}[1] {\vspace{12pt}
        \refstepcounter{appendixc}
        \setcounter{figure}{0}
        \setcounter{table}{0}
        \setcounter{lemma}{0}
        \setcounter{theorem}{0}
        \setcounter{corollary}{0}
        \setcounter{definition}{0}
        \setcounter{equation}{0}
        \renewcommand{\thefigure}{\Alph{appendixc}.\arabic{figure}}
        \renewcommand{\thetable}{\Alph{appendixc}.\arabic{table}}
        \renewcommand{\theappendixc}{\Alph{appendixc}}
        \renewcommand{\thelemma}{\Alph{appendixc}.\arabic{lemma}}
        \renewcommand{\thetheorem}{\Alph{appendixc}.\arabic{theorem}}
        \renewcommand{\thedefinition}{\Alph{appendixc}.\arabic{definition}}
        \renewcommand{\thecorollary}{\Alph{appendixc}.\arabic{corollary}}
        \renewcommand{\theequation}{\Alph{appendixc}.\arabic{equation}}
        \noindent{\tenbf Appendix \theappendixc #1}\par\vspace{5pt}}
\newcommand{\subappendix}[1] {\vspace{12pt}
        \refstepcounter{subappendixc}
        \noindent{\bf Appendix \thesubappendixc. {\kern1pt \bfit #1}}
        \par\vspace{5pt}}
\newcommand{\subsubappendix}[1] {\vspace{12pt}
        \refstepcounter{subsubappendixc}
        \noindent{\rm Appendix \thesubsubappendixc. {\kern1pt \tenit #1}}
        \par\vspace{5pt}}

\newcommand{\textlineskip}{\baselineskip=13pt}
\newcommand{\smalllineskip}{\baselineskip=10pt}

\def\eightcirc{
\begin{picture}(0,0)
\put(4.4,1.8){\circle{6.5}}
\end{picture}}
\def\eightcopyright{\eightcirc\kern2.7pt\hbox{\eightrm c}}

\newcommand{\publisher}[2]{{\begin{center}\footnotesize\smalllineskip
        Received #1\\
        Revised #2
        \end{center}
        }}

\def\abstracts#1#2#3{{
        \centering{\begin{minipage}{4.5in}\baselineskip=10pt\footnotesize
        \parindent=0pt #1\par
        \parindent=15pt #2\par
        \parindent=15pt #3
        \end{minipage}}\par}}

\renewenvironment{thebibliography}[1]                   
        {\frenchspacing
         \ninerm\baselineskip=11pt
         \begin{list}{\arabic{enumi}.}
        {\usecounter{enumi}\setlength{\parsep}{0pt}
         \setlength{\leftmargin 12.7pt}{\rightmargin 0pt} 
         \setlength{\itemsep}{0pt} \settowidth
        {\labelwidth}{#1.}\sloppy}}{\end{list}}

\newcounter{itemlistc}
\newcounter{romanlistc}
\newcounter{alphlistc}
\newcounter{arabiclistc}

\newenvironment{romanlist}
        {\setcounter{romanlistc}{0}
         \begin{list}{$($\roman{romanlistc}$)$}
        {\usecounter{romanlistc}
         \setlength{\parsep}{0pt}
         \setlength{\itemsep}{0pt}}}{\end{list}}

\newcommand{\fcaption}[1]{
        \refstepcounter{figure}
        \setbox\@tempboxa = \hbox{\footnotesize Fig.~\thefigure. #1}
        \ifdim \wd\@tempboxa > 5in
           {\begin{center}
        \parbox{5in}{\footnotesize\smalllineskip Fig.~\thefigure. #1}
            \end{center}}
        \else
             {\begin{center}
             {\footnotesize Fig.~\thefigure. #1}
              \end{center}}
        \fi}

\newcommand{\tcaption}[1]{
        \refstepcounter{table}
        \setbox\@tempboxa = \hbox{\footnotesize Table~\thetable. #1}
        \ifdim \wd\@tempboxa > 5in
           {\begin{center}
        \parbox{5in}{\footnotesize\smalllineskip Table~\thetable. #1}
            \end{center}}
        \else
             {\begin{center}
             {\footnotesize Table~\thetable. #1}
              \end{center}}
        \fi}

\def\@citex[#1]#2{\if@filesw\immediate\write\@auxout
        {\string\citation{#2}}\fi
\def\@citea{}\@cite{\@for\@citeb:=#2\do
        {\@citea\def\@citea{,}\@ifundefined
        {b@\@citeb}{{\bf ?}\@warning
        {Citation `\@citeb' on page \thepage \space undefined}}
        {\csname b@\@citeb\endcsname}}}{#1}}

\newif\if@cghi
\def\cite{\@cghitrue\@ifnextchar [{\@tempswatrue
        \@citex}{\@tempswafalse\@citex[]}}
\def\citelow{\@cghifalse\@ifnextchar [{\@tempswatrue
        \@citex}{\@tempswafalse\@citex[]}}
\def\@cite#1#2{{$\null^{#1}$\if@tempswa\typeout
        {IJCGA warning: optional citation argument
        ignored: `#2'} \fi}}

\def\pmb#1{\setbox0=\hbox{#1}
        \kern-.025em\copy0\kern-\wd0
        \kern.05em\copy0\kern-\wd0
        \kern-.025em\raise.0433em\box0}

\def\fnt#1#2{\footnotetext{\kern-.3em
        {$^{\mbox{\scriptsize #1}}$}{#2}}}

\def\fpage#1{\begingroup
\voffset=.3in
\thispagestyle{empty}\begin{table}[b]\centerline{\footnotesize #1}
        \end{table}\endgroup}

\def\runninghead#1#2{\pagestyle{myheadings}
\markboth{{\protect\footnotesize\it{\quad #1}}\hfill}
{\hfill{\protect\footnotesize\it{#2\quad}}}}
\font\tenrm=cmr10
\font\tenit=cmti10
\font\tenbf=cmbx10
\font\bfit=cmbxti10 at 10pt
\font\ninerm=cmr9

\font\eightrm=cmr8

\def\qed{\hbox{${\vcenter{\vbox{                        
   \hrule height 0.4pt\hbox{\vrule width 0.4pt height 6pt
   \kern5pt\vrule width 0.4pt}\hrule height 0.4pt}}}$}}

\renewcommand{\thefootnote}{\fnsymbol{footnote}}        

\def\bsc{{\sc a\kern-6.4pt\sc a\kern-6.4pt\sc a}}       
\def\bflatex{\bf L\kern-.30em\raise.3ex\hbox{\bsc}\kern-.14em
T\kern-.1667em\lower.7ex\hbox{E}\kern-.125em X}

\begin{document}

\runninghead{Loop models, Marginally Rough Interfaces, and 
the Coulomb Gas} 
{Loop models, Margianlly Rough Interfaces, and the Coulomb Gas}

\normalsize\textlineskip
\thispagestyle{empty}
\setcounter{page}{1}



\fpage{1}
\centerline{\bf LOOP MODELS, MARGINALLY ROUGH INTERFACES, }
\vspace*{0.035truein}
\centerline{\bf AND THE COULOMB GAS
\footnote{Parts of this work were done in collaboration 
with C.L. Henley, J. deGier, and B. Nienhuis.}}
\vspace*{0.37truein}
\centerline{\footnotesize JAN\'E KONDEV}
\vspace*{0.015truein}
\centerline{\footnotesize\it Department of Physics, Brown University}
\baselineskip=10pt
\centerline{\footnotesize\it Providence, Rhode Island, O2912-1843, USA}
\vspace*{0.225truein}
\publisher{(received date)}{(revised date)}

\vspace*{0.21truein}
\abstracts{We develop a  coarse-graining procedure for two-dimensional 
models of fluctuating loops 
by mapping them to interface models. The
result  is an
effective field theory for the scaling limit of loop models, which is
found to be a Liouville theory with imaginary couplings. This 
field theory is {\it completely specified} by geometry and conformal
invariance  {\it alone}, and it  leads to exact results for the 
critical  exponents and the 
conformal charge of loop models. A  physical interpretation of the 
Dotsenko-Fateev screening charge is found. 
}{}{}

\input epsf


\vspace*{1pt}\textlineskip      
\section{Introduction}    
\vspace*{-0.5pt}

\textheight=7.8truein
\setcounter{footnote}{0}
\renewcommand{\thefootnote}{\alph{footnote}}
\noindent
Loop models are defined by drawing closed  loops (which can come in 
one or more flavors)
along the bonds of a two-dimensional lattice ${\cal L}$, subject to the 
constraint that only loops of different flavor may cross. 
Boltzmann weights of different loop configurations are
completely determined by 
specifying the weight ($\rho$) of all possible loop arrangements at a 
single vertex of ${\cal L}$, and the fugacity ($n$) 
of every loop flavor. 
The partition function can be written as:~\cite{warn}
\begin{equation}
\label{part}
  Z = \sum_{\cal G} \rho_1^{m_1} \rho_2^{m_2} \ldots \rho_V^{m_V} n_1^{l_1}
  n_2^{l_2} \ldots n_F^{l_F} \; ,
\end{equation} 
where $F$ is the number of loop flavors and  $V$ the 
number of allowed vertex  configurations.
The sum in Eq.~(\ref{part}) goes  over 
all allowed loop configurations ${\cal G}$; $m_i$ ($i=1,\ldots,V$) 
is the number of 
vertices of type $i$, and  $l_j$ ($j=1,\ldots,F$) 
the number of loops of flavor $j$ in ${\cal G}$.

The  motivation for studying 
loop models comes from the observation that they 
are rather ubiquitous in the realm of two-dimensional statistical 
mechanics. Loop models
appear very naturally when one considers domain boundaries
in discrete spin models. For example, 
model $A$ in Fig.1  is equivalent to the 
$Q=n^2$ Potts model on the square lattice, where the loops 
run along the bonds of the medial lattice so as to encircle 
clusters of equal Potts spin.~\cite{baxterbook} 
In a somewhat different 
setting, loop models appear as space-time diagrams of certain 
one-dimensional quantum
systems, where the quantum partition function can be written as a
sum over loop configurations with appropriate weights. 
This {\em loop representation} of quantum models 
has been successfully taken advantage of in recent 
Monte Carlo (MC) simulations of spin chains and ladders.\cite{evertz}
The loop representation allows for non-local (loop) update moves
in the MC algorithm,  
which lead to a considerable reduction in critical slowing down,
and rather large system sizes can be simulated.  
 On the  analytic side, the  loop representation of the  
Heisenberg spin chain  
has  lead to a very nice geometrical interpretation  of 
correlation functions.\cite{aizenman} For instance,   the 
spin-spin correlation function is associated with the 
probability that two
points on the lattice are connected by a loop segment,  
in the appropriate loop model.

\begin{figure}[htbp]
\label{fig}
\vspace*{13pt}
\epsfxsize=12.5cm \epsfbox{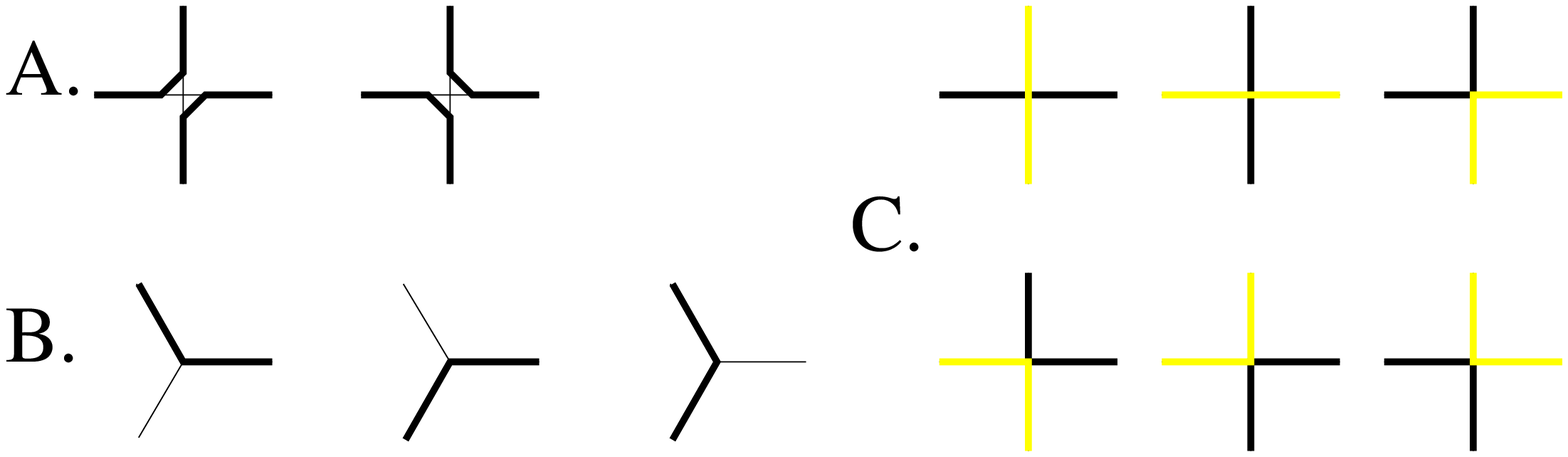}
\vspace*{13pt}
\fcaption{Allowed vertex configurations  in  fully packed loop models 
on the square and  honeycomb lattice; every vertex is assigned the 
weight $\rho=1$.  The models $A$ and $B$ have 
a single loop flavor with fugacity $n$, while  loops in model $C$
come in two flavors (black and grey), with fugacities $n_1$ and $n_2$.}  
\end{figure}

For a specific choice of Boltzmann weights, loop models can be in a 
critical state characterized by power law correlations.  
Here we  examine the critical properties of fully packed loop models.
 These loop  models  have 
non-vanishing and identical ($\rho_i=1$) vertex weights only for 
configurations 
that satisfy the fully-packing constraint; this constraint ensures that 
every vertex of the lattice is covered by one, and only one 
loop of each flavor. We have studied three such models, shown in 
Fig.1. Model $A$ (No.\ of loop flavors, $F=1$) 
is equivalent to the loop  
decomposition of the critical $Q=n^2$ Potts model,\cite{baxterbook} 
model $B$ ($F=1$) is the zero-temperature phase 
of the $O(n)$ model,~\cite{nienblote,batch,jknien} and 
model $C$ ($F=2$)
has recently been introduced as the loop generalization of the 
four-coloring model on the square lattice.~\cite{jkprb} The theoretical 
challenge we are faced with, is
to find an effective description of a critical loop model in terms 
of a (conformal) field theory, 
from which critical exponents and other universal quantities can
be calculated exactly. 

The answer to the above stated theoretical challenge is provided by 
mapping a loop model to an {\em interface model}, where loops are
identified with contour lines of the interface. Coarse-graining of
the interface model  leads  to  
an effective field theory which describes the 
critical fluctuations of loops
in terms of a fluctuating height field.  
This effective field theory is equivalent by a duality to a 
Cou\-lomb gas of electric and magnetic charges, with an 
additional electric charge placed at infinity. 
%
%

In the Coulomb gas approach to two-dimensional critical phenomena, 
the critical phase is identified with the vacuum phase of a 
gas of electric and magnetic charges;\cite{nienrev} 
the strength of the Coulomb
interaction is parameterized by the coupling constant $g$. 
Critical exponents 
are readily calculated  once the value of this coupling is known. 
In the interface representation of loop models, $g$ 
has the physical interpretation
of the effective surface tension, 
or stiffness (see Eq.~(\ref{action})). 
The effective stiffness is not a universal quantity and 
to calculate it in terms of the microscopic weights 
is typically as hard as solving the model 
exactly.\footnote{In the past, the   
value of the coupling constant has been determined for models that map 
to exactly solvable models by comparing the Coulomb-gas expression for
a particular exponent, which is $g$ dependent,
with the value of the same exponent found from the 
exact solution.\cite{nienrev}}    
In this paper we  show that the stiffness of the interface 
is {\it completely specified} 
by geometry and conformal invariance {\it alone}. 
More precisely, it is shown that demanding conformal invariance of the 
effective
field theory of critical loop models completely fixes all 
the couplings  in the theory.

\section{Interface representations of loop models}
\noindent
The {\it general} procedure for constructing an effective field 
theory of a loop model is outlined here for the $A$ model. Details 
of this construction for models $B$ and $C$ can be found 
elsewhere.\cite{jknien,jkunp}

Model $A$ is the well known polygon representation 
of the $Q=n^2$ Potts model, and it is critical for 
$0\le n\le 2$.\cite{baxterbook} 
For $n>2$ the model has a finite correlation 
length, roughly the size of the largest loop; in the limit 
$n\rightarrow \infty$
it orders  in one of two states, which are related by a lattice translation,
and in which all the loops are of length four.

The construction of an effective field theory for the 
scaling limit of a loop  model can be broken up into three steps:

\subsection{Orient the loops}
\noindent
Each loop is assigned an orientation by placing arrows along the
bonds of the square lattice. In order to assign proper weights to 
the loops ($n\le2$) every  left (right) turn is weighted by the 
phase factor $\exp(-{\rm i}e_0 \pi/4)$ 
($\exp({\rm i}e_0 \pi/4)$).\cite{nienrev} For
every closed loop on the 
square lattice the difference between the number of right, and the 
number of left turns is $\pm4$. Therefore, by summing
over the two possible orientations each loop is assigned the weight 
\begin{equation}
\label{fuge0}
 n = 2 \cos(\pi e_0) \; . 
\end{equation}

\subsection{Map to an interface model}
\noindent
Each loop is viewed as a {\it contour line}
of an interface model. For the $A$ model this is the 
well known body-centered solid-on-solid (BCSOS) model;\cite{vanb}
for the  $B$ and $C$ models, interface 
models with  two and three component 
height variables are obtained.\cite{huse,jkprb,jknpb} 
The microscopic heights $z$,  of model $A$, are defined 
on the dual lattice, and $z$ increases (decreases) by 
$1/2$ every time an oriented loop is crossed from left to right 
(right to left). Coarse-graining of the microscopic height
leads to an effective field theory of the loop model.  

\subsection{Coarse grain the interface model}
\noindent
The interface model is coarse-grained in 
terms of the {\it ideal states}.
These states minimize the variance of the 
microscopic height, i.e., they are
macroscopically flat. In the loop model these are states 
with the maximum
number of loops on the lattice, or equivalently,  the states which 
one finds in the limit $n\rightarrow \infty$.  
For finite $n$ the lattice breaks up into domains of ideal states,
and each domain is assigned a {\it coarse-grained height} equal to the 
average microscopic height over the domain: $h=<\!z\!>$. 
The continuum limit is taken by replacing the discrete heights  
by a height field $h({\bf x})$.  
The height field $h({\bf x})$ is compactified
to a circle, i.e.,   
\begin{equation}
\label{compact}
h \equiv h + {\cal R} \; , \; \; {\cal R}=\{0,\pm1,\pm2,\ldots\} 
\end{equation}
where ${\cal R}$ is the {\it repeat} lattice; 
the vectors in this lattice
are height differences between equivalent loop configurations. 

The Euclidean action of  the effective field theory,  which 
describes the long-wave\-length fluctuations of the microscopic height
(or the critical fluctuations of the loops in the $A$ model),  
can be written  in terms of the 
field $\phi({\bf x})\equiv 2\pi h({\bf x})$ as
\begin{eqnarray}
\label{action}
 S[\phi] & = & S_g[\phi] + S_b[\phi] + S_p[\phi] \nonumber \\ 
 S_g  & = & \frac{g}{4\pi} \int \!\! d^2 {\bf x} \;
                                             [\nabla \phi]^2 \nonumber \\
 S_b  & = & \frac{{\rm i} e_0}{4 \pi} \int \!\! d^2 {\bf x} \; R \: \phi 
                                                              \nonumber \\
 S_p  & = & \int \!\! d^2 {\bf x} \; w(\phi) 
\end{eqnarray}

The three parts to the action have a simple geometrical interpretation:

\begin{romanlist}
\item 
$S_g$ describes the Gaussian fluctuations of the height 
around the flat ideal states. 

\item
$S_b$ is  the coupling of  the height to the scalar 
curvature $R$. Namely, if a loop winds around a point 
of non-zero curvature, 
then the
difference between the number of 
left and the number of right turns, for this 
loop, is no longer four; $S_b$ corrects for this. 
For the square lattice in 
the infinite plane this term corresponds to a 
{\it background charge} $2e_0$ placed at the point at 
infinity. 

\item
 $S_p$ is the potential term. Its origins are 
twofold: 1) it
accounts for the discrete nature of the heights, and 2) 
it assigns proper weights 
($n<2$) to the loops, i.e., the operator $w(\phi)$ is conjugate to 
the loop weight. From Eq.(\ref{compact}) we  conclude that 
$w(\phi)$ can be expanded into a Fourier series, 
\begin{equation}
\label{fseries}
w(\phi) = \sum_{E \in \cal{R}^\ast}  w_E \; e^{{\rm i} E \phi} \; , 
\end{equation}
where ${\cal R}^\ast$ is the lattice {\it dual} to ${\cal R}$; here 
${\cal R}^\ast$
is simply the set of integers.\footnote{Elements
of ${\cal R}^\ast$ are the electric charges, while elements of  
${\cal R}$ are the magnetic charges of the 
two-dimensional Coulomb gas, which is dual to the interface model.} 
Only the most relevant {\it vertex operator} (i.e., exponential of 
the height field)
appearing in 
the above Fourier expansion is kept in the action.
Upon examining the values which $w(\phi)$ takes in the ideal states,
we find this to be the  
vertex operator with {\it electric} charge 
\begin{equation}
\label{charge}
E_w = 2 \; .
\end{equation}
\end{romanlist}

\section{Calculation of the coupling}
\noindent
In order to calculate the value of the coupling $g$, for model 
$A$ in the critical phase ($0\le n\le 2$), we
demand that the action $S$ describe a conformally 
invariant field theory. This, as will be made clear below, 
is equivalent to the assumption that the loop fugacity does 
not change with the change of scale.   

For $S$ to describe a conformal field theory the potential term 
has to be {\it marginal}, i.e., the scaling dimension 
$x_w$ of the operator $w(\phi)$ is two.\cite{dots2+mathur} 
The dimension $x_w$ can be expressed in terms of the coupling $g$ 
and the background charge $e_0$ as:\cite{dots} 
\begin{equation}
\label{dim}
x_w = E_w (E_w-2e_0) / 2g = 2 \; .
\end{equation}
Using Eq.(\ref{charge}) we find  the {\it exact} value of the 
coupling
\begin{equation}
\label{exactg}
 g = 1 - e_0  \; ; 
\end{equation}
both the background charge and $g$ are continuous functions of the 
fugacity $n$, Eq.(\ref{fuge0}). 
This value of the coupling agrees with the value found  from the exact
solution of the eight-vertex model.\cite{nienrev} We emphasize that
$e_0(n)$ in Eq.(\ref{fuge0}) defines a whole {\em family}
of critical models, whose conformal charges are given in Eq.(\ref{ccharge}).  

\section{Discussion}
\noindent
The field theory described by the Euclidean action in Eq.~(\ref{action})
is a Liouville theory with imaginary couplings, which has been 
suggested as the Lagrangean description of a two-dimensional 
Coulomb gas in the presence of a background charge.\cite{dots2+mathur} 
The potential term $S_p$ is the {\it screening charge},
which was originally 
introduced into the Coulomb gas to ensure that the four-point
correlation functions do not vanish.\cite{dots} Here this operator 
appears quite naturally in the action as a result of 
the coarse graining, and 
it is the operator conjugate to the loop fugacity.

The term $S_p$ also contains 
the {\em locking potential}, 
which enforces the condition that the heights 
are discrete. We find that the locking potential is {\it marginal} 
along the whole critical 
line ($n\le2$) of the loop model. This leads to the 
somewhat surprising
conclusion that the associated interface model is at the 
{\it roughening transition} for {\em all} $n\le2$, 
not just at the boundary at $n=2$. In the case of model $B$ Baxter
arrived at the same conclusion from the exact solution of the 
related three-coloring model; he showed that the partition function
has a line of essential singularities for 
$0\le n\le 2$.\cite{baxter70,jknien}   
 
For the $B$ and $C$ models the height field has more 
then one component and consequently the magnetic 
(${\bf M}$) and  electric charges (${\bf E}$)
are {\it lattice vectors} in  ${\cal R}$, 
and the dual ${\cal R}^\ast$.
The proper identification of these lattices follows 
from the coarse-graining procedure.
The fact that the height is {\it compactified} on ${\cal R}$, 
Eq.(\ref{compact}),  
leads to some interesting conclusions about the symmetry of 
these loop models  in the continuum.  
Namely, for  $n=2$ ($B$ model) and $n_1=n_2=2$ ($C$ model) the 
background
charge vanishes, and these loop models are described, 
in the continuum, by 
the $SU(3)_{k=1}$ and the $SU(4)_{k=1}$ Wess-Zumino-Witten model 
respectively;\cite{read,jknpb} for the $A$ model,
at $n=2$ one finds the $SU(2)_{k=1}$ WZW model.\cite{aff,jknpb} 
The action written in terms of the
height field is the free-field representation of these sigma models.

The conformal charge and all the critical exponents of the loop 
models can 
be expressed in terms of the coupling $g$ and the background charge 
${\bf e}_0$:\cite{dots}
\begin{eqnarray}
\label{ccharge}
c & = &  k - \frac{6{\bf e}_0^2}{g}  \nonumber \\
x({\bf E}, {\bf M}) & = & \frac{1}{2 g} \: 
  {\bf E}\cdot({\bf E}-2{\bf e}_0)
 + \frac{g}{2} \: {\bf M}^2 
\; ,
\end{eqnarray} 
where $x({\bf E}, {\bf M})$ is the scaling dimension of 
the operator with electro-magnetic charge 
$({\bf E}, {\bf M})$, and $k=1,2$, and $3$ for the $A$,$B$,
and $C$ model respectively. 
For example, the m-RSOS models 
which are described in the 
continuum by the minimal models of conformal field theory 
can be mapped to the the $A$ model with 
$e_0 = 1/(m+1), (m>2)$.\cite{nienfoda} 
The conformal charge that we calculate from 
Eqs.~(\ref{exactg}) and~(\ref{ccharge}) is $ c = 1 - 6/m(m+1)$, 
which is the well known expression.

\nonumsection{Acknowledgments}
It is a pleasure to acknowledge illuminating discussions with 
T. Spencer, J.B. Marston, and in particular J. Cardy, 
who pointed out to me the connection between marginal operators and
screening charges in the Coulomb gas.
I am indebted to C.L. Henley, J. deGier, and B. Nienhuis, my 
collaborators on related projects, who have helped shape my 
understanding of loop models in a significant way. 
This work was supported by the NSF through grant No. DMR 9357613. 

\nonumsection{References}


\noindent
\begin{thebibliography}{000}
\bibitem{warn}
   S.O. Warnaar and B. Nienhuis, J. Phys. A {\bf 26}, 2301 (1993).
\bibitem{baxterbook}
   R.J. Baxter, {\em Exactly Solved Models in Statistical Mechanics}
   (Academic Press, 1982)
\bibitem{evertz}
   H.G. Evertz, G. Lana, and M. Marcu, Phys. Rev. Lett. {\bf 70},
   875 (1993); N. Kawashima and J.E. Gubernatis, Phys. Rev. Lett. 
   {\bf 73}, 1295 (1994); B. Frischmut, B. Ammon, and M. Troyer, 
   preprint No. cond-mat/9601025. 
\bibitem{aizenman}
   M. Aizenman and B. Nachtergaele, Comm. Math. Phys. {\bf 164}, 17 
     (1994). 
\bibitem{nienblote}
   H.W.J. Bl\"ote and B. Nienhuis, Phys. Rev. Lett. {\bf 72}, 1372
    (1994). 
\bibitem{batch}
   M.T. Batchelor, J. Suzuki, and C.M. Yung, Phys. Rev. Lett. {\bf 73},
   2646 (1994).
\bibitem{jknien}
    J. Kondev, J. deGier, and B. Nienhuis, submitted to {\it 
        J. Phys. A: Math. Gen.}; preprint No. cond-mat/9603170.
\bibitem{jkprb}
    J. Kondev and C.L. Henley, Phys. Rev. {\bf B52}, 6628 (1995).
\bibitem{nienrev}
    B. Nienhuis, in {\em Phase Transitions and Critical Phenomena},
     Vol.11, ed. C. Domb and J.L.  Lebowitz (Academic Press, New York,
     1987) 
\bibitem{jknpb}
    J. Kondev and C.L. Henley, Nucl. Phys. {\bf B464}, 540 (1996).
\bibitem{jkunp}
    J. Kondev, in preparation.
\bibitem{vanb}
    H. van Beijeren, Phys. Rev. Lett. {\bf 38}, 993 (1977).
\bibitem{huse}
    D.A. Huse and A.D. Rutenberg, Phys. Rev. {\bf B45}, 7536 (1992).
\bibitem{dots2+mathur}
    Vl.S. Dotsenko and V.A. Fateev, Nucl. Phys. {\bf B251}, 691 (1985);
    S.D. Mathur, Nucl. Phys. {\bf B369}, 433 (1992).
\bibitem{dots}
    Vl.S. Dotsenko and V.A. Fateev, Nucl. Phys. {\bf B240}, 312 (1984).
\bibitem{baxter70} 
    R.J. Baxter, J. Math. Phys. {\bf 11}, 784 (1970). 
\bibitem{read}
    N. Read, reported in Kagom\'e workshop (Jan. 1992), unpublished.
\bibitem{aff} 
    I. Affleck, Phys. Rev. Lett. {\bf 55}, 1355 (1985).
\bibitem{nienfoda}
    O. Foda and B. Nienhuis, Nucl. Phys. {\bf B324}, 643 (1989). 

\end{thebibliography}
\end{document}